\documentclass[12pt]{article}

\usepackage{fullpage}
\usepackage[utf8]{inputenc}
\usepackage{setspace}
\usepackage{parskip}
\usepackage{titlesec}
\usepackage[section]{placeins}
\usepackage{xcolor}
\usepackage{breakcites}
\usepackage{lineno}
\usepackage{hyphenat}
\usepackage{times}

\titlespacing{\section}{0pt}{*3}{*1}
\titlespacing{\subsection}{0pt}{*2}{*0.5}
\titlespacing{\subsubsection}{0pt}{*1.5}{0pt}

\usepackage{authblk}

\usepackage{graphicx}
\usepackage[space]{grffile}
\usepackage{latexsym}
\usepackage{textcomp}
\usepackage{longtable}
\usepackage{tabulary}
\usepackage{booktabs,array,multirow}
\usepackage{amsfonts,amsmath,amssymb}
\providecommand\citet{\cite}
\providecommand\citep{\cite}

\usepackage[utf8]{inputenc}
\usepackage[english]{babel}

\begin{document}

\title{Postseismicity of slow-slip doublets discerned on the outermost of the Nankai Trough subduction megathrust}

\author[1*]{Dye SK Sato}%
\author[1]{Takane Hori}%
\author[1]{Takeshi Iinuma}%
\author[2]{Masayuki Kano}%
\author[3]{Yusuke Tanaka}%
\affil[1]{Research Institute for Marine Geodynamics, Japan Agency for Marine-Earth Science and Technology, 
    Yokohama 236‑0001, Japan}%
\affil[2]{Disaster Prevention Research Institute, Kyoto University, Uji 611-0011, Japan}%
\affil[3]{Geography and Crustal Dynamics Research Center, Geospatial Information Authority of Japan, Tsukuba 305-0811, Japan}%

\vspace{-1em}

  \date{November 25, 2025}


\begingroup
\let\center\flushleft
\let\endcenter\endflushleft
\maketitle
\endgroup

\doublespacing

\selectlanguage{english}
{\bf Despite dissimilar slip rates, slow earthquakes are faulting as ordinary earthquakes are. It is therefore physically natural that slow earthquakes also cause postseismic motions similarly to ordinary earthquakes, even though coseismic and postseismic slips remain undifferentiated for slow earthquakes. We pursue the slow-earthquake postseismicity based on the analysis of a fault slip beneath the Bungo Channel, the westernmost region of the Nankai Trough subduction zone in southwestern Japan. Its 2010 long-term slow slip event (SSE) was mispredicted by physics-based models, which concludes that the initial acceleration of this SSE was too abrupt for a slow variant of a fault rupture. We identify that a mispredicted GNSS signal evolves logarithmically in time, preceded by minor signals that evolve exponentially, lasting about two years west and about half a year east. By performing sparse inverse modeling on the GNSS, we have estimated that exponential slips occur at the same depth, bracketing a logarithmic slip that occurs beneath the channel. The regions of exponential slips match repeating slow-slip regions, and deep tremors synchronize exclusively with the logarithmic slip. This source complexity can be explained as a neighboring rupture doublet and its afterslip and aftershocks by the known mechanics of ordinary earthquakes. If slow earthquakes have a dual origin in exponentially nucleating slow rupture and logarithmically decelerating postseismic creep, it is possible to pick the slow earthquake nuclei that could accelerate into megathrust catastrophes.\\}

Slow earthquakes have been discovered as a ubiquitous mode of crustal stress release, which occurs intermittently as ordinary earthquakes but with slow moment-release rates atypical of earthquakes~\cite{hirose1999slow,obara2002nonvolcanic,miller2002periodic,rogers2003episodic}. This class of phenomena spatially separates from regular earthquake sources~\cite{yagi2003partitioning,kato2012propagation} and is sometimes thought to precede megathrust earthquakes~\cite{kanamori1974focal,kato2012propagation,ito2013episodic,bletery2023precursory}. 
Whereas a large portion of their nature remains enigmatic, evidence is accumulating that slow earthquakes are faulting, namely fault slip faster than the plate loading rate as ordinary earthquakes.
The observation of the slow-earthquake occurrence on the plate interface as shear slip, isolated from the ordinary earthquakes, supports the interpretation that these sporadic slow events are strain release phenomena on plate boundaries~\cite{kawasaki2001space,ozawa2002detection,shelly2006low,wech2007cascadia,ide2007mechanism,nishikawa2019slow}. 

Faulting takes two types: spontaneous and triggered. 
In a spontaneous event, fault rupture occurs to release the load stress built up on the locked zone of the interface~\cite{kostrov1988principles}. 
Spontaneous rupture repeats in a seismic cycle, which contains the interseismic phase where stress is loaded, the coseismic phase where the locked zone slips, and the postseismic phase which causes afterslip~\cite{avouac2015geodetic}. The afterslip refers to the residue of stable creeping in the surrounding area that follows the coseismic acceleration of the rupture zone. While the zone subjected to coseismic rupture quickly returns to a locked state from steady-state sliding at kinetic friction, the creeping zone stably continues to gradually decelerate until the accumulated moment in the stress shadow is removed~\cite{perfettini2004postseismic,kato2004interaction,kato2007expansion}. 
In a triggered event, the zone at the earthquake preparatory stage responds to perturbation and hosts a rupture event. 
The surrounding area hosts decelerating stable creep for both spontaneous and triggered ruptures, as observed in large foreshock-mainshock sequences~\cite{helmstetter2009afterslip,kato20162016}. 

Geodetic analyses of the slow slip event (SSE) have reported that these two types of faulting are valid also for slow earthquakes~\cite{cornet1997seismic,miller2002periodic,peng2009remote,materna2019dynamically}. 
There are repeaters of SSEs similar to those of ordinary earthquakes~\cite{miller2002periodic,ozawa2003characteristic}, which evidence slow-earthquake cycles that vary plate coupling over time and spontaneously cause ruptures~\cite{materna2019dynamically,bartlow2020long,wallace2020slow}. 
Aseismic slips induced by fluid injection~\cite{cornet1997seismic,bhattacharya2019fluid,guglielmi2015seismicity} and teleseismic waves~\cite{peng2009remote} exemplify the triggered slow earthquakes. 

If faulting entails postseismic slips regardless of faulting types and slow earthquakes are faulting, it is mechanically natural that slow earthquakes also define postseismicity.
However, this mechanical expectation is not yet supported by observed data. 
Even theoretical models have not explicitly addressed the postseismic phenomena that follow slow earthquakes, despite the empirical knowledge from earthquake sequence simulations employing artificial damping that coseismic slip rates have little effect on the occurrence of postseismic creeps~\cite{rice1993spatio}.
Presumably, the term `slow earthquake' implicitly equates the two mechanically distinct fault motions, coseismic rupture and postseismic creep, in the kinematic sense of being slow. Descriptions of slow earthquakes as slow coseismic slips are being challenged by misprediction analyses in fact, where the difference between observed data and expectations from laboratory-derived physics is quantified~\cite{kano2024data,fukushima2025spatio}. 
This study reports an example of SSEs where the observed behavior is naturally understood if the postseismic signal of the slow earthquake has remained unrecognized but present.

\section*{Analysis}
\subsection*{Hypothetical signals of slow-slip's afterslip}

\begin{figure} 
	\includegraphics[width=1.0\textwidth]{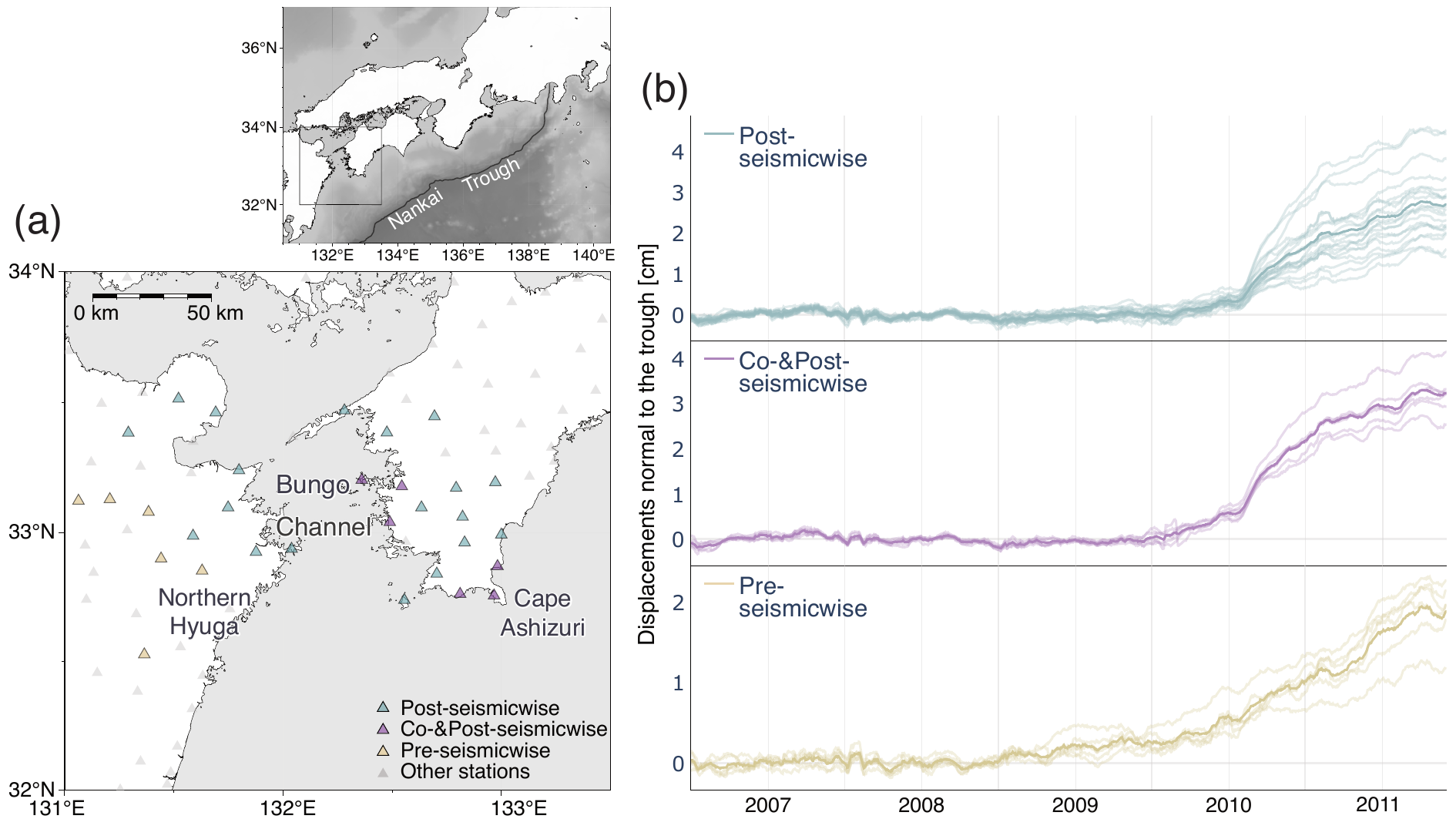}
  \caption{Study area and data set. (a) 
  Selected GNSS stations surrounding the Bungo Channel. 83 stations were selected from 86 stations located within 131--134$^{\circ}$E and 32--34$^{\circ}$N, excluding three outliers where annual variations exceed 0.5 cm. Three groups of distinctive trends are colored blue, purple, and khaki, while the others are shown in gray. Geographical names appearing in the text are also indicated. 
  (b) Three characteristic signal trends of the 2010 Bungo Channel SSE. The horizontal components of GNSS displacements normal to the trough are plotted from July 2006 to June 2011 in light colors, with their stacked values in dark colors. The color scheme follows Fig.~\ref{fig:1}a. 
  Data were processed to remove signals of antenna replacements and linear trends fitted within a pre-event window from 2006.5 to 2008.5, filtered using a moving average over one month subsequent to the date on the horizontal axis.
}
  \label{fig:1}
\end{figure}

The Bungo Channel is situated at the westernmost of the Nankai subduction zone in southwestern Japan (Fig.~\ref{fig:1}a), which recurrently hosts megathrust earthquakes with moment magnitudes approximately 8 or grater~\cite{furumura2011revised}. Long-term SSEs repeat every 6--8 years in this outermost area of the megathrust~\cite{hirose2023long}. 
The 2010 SSE is a closely examined event, where the slip begins in September 2009 near the north of Cape Ashizuri, apparently migrates to the Bungo Channel around February 2010, and lessens around September 2010~\cite{hirose2010source,ozawa2013spatial,yoshioka2015spatiotemporal}.
Data-assimilation analyses based on rate-and-state friction suggest that the strong signal observed in the Bungo Channel~\cite{hirose2010source,ozawa2013spatial} was too abrupt to be considered a slow rupture event of a stress-loaded locked zone~\cite{kano2024data,fukushima2025spatio}.

Figure~\ref{fig:1}a shows the 83 GNSS stations located within 131--134$^{\circ}$E and 32--34$^{\circ}$N, excluding three outliers where annual variations of recorded displacements exceed 0.5 cm. 
For better visibility of the time series, the stations are colored into groups according to the trend of their time evolution. 
The grouping was subjectively made to reflect the three qualitative trends observed in the daily coordinates of the GNSS~(Fig.~\ref{fig:1}b). 
Figure~\ref{fig:1}b shows the detrended trough-normal components of daily coordinates in the period from July 2006 to June 2011, or specifically, [2006-07-01, 2011-06-01]. 
Lines of light colors represent the set of time series for stations of the same color, and dark colors represent the stacked data.
Linear detrending has been made by least-squares to fit the pre-event time window from 2006.5 to 2008.5~\cite{kano2024data}. One-month moving averages are also taken to suppress short-term noise dominated by tidal effects. 
Along-strike data were similarly processed for the later appearing inverse analysis.

We observe the following three characteristic trends in the grouped signals. 
Figure~\ref{fig:1}b blue captures the blow-up signal starting from the onset of the SSE right beneath the Bungo Channel, which was mispredicted by physics-based models as a smooth time evolution [Fig. 1b in~Ref.\cite{kano2024data}]. Figure~\ref{fig:1}b purple captures a preceding weak signal along with the blow-up signal. This preceding run-up corresponds to the initial episode of the SSE~\cite{ozawa2013spatial} that began east of the Channel in mid-2009 prior to the active slip beneath the channel. Figure~\ref{fig:1}b khaki captures another gradual signal lasting for about two years prior to that active slip, although such a signal has been unmodeled in the previous analyses of this SSE.

Given their appearances, we refer to them as postseismicwise, co-\&post-seismicwise, and preseismicwise groups, respectively. 
The associated trends and signals are termed accordingly. 
The blow-up trend captured in the postseismicwise group (Fig.~\ref{fig:1}b, blue line) shares the same onset as the slow slip right beneath the Bungo Channel starting in February 2010~\cite{hirose2010source}, suggesting that this blow-up traces the slip evolution beneath the channel. Data assimilation analysis using rate-and-state faults~\cite{kano2024data,fukushima2025spatio} implied that the rapid acceleration initiating this blow-up is unpredicted within the physics of fault rupture. 
The postseismicwise group shown in Fig.~\ref{fig:1}b indicates that the blow-up, not expected as a fault rupture, follows a logarithm typical of afterslip. In the postseismic phase, fault motion is a stable sliding, termed creep, that slowly continues to decelerate toward a steady state; the resulting postseismic slip logarithmically increases, with the slip rate inversely proportional to time consistent with the Omori law~\cite{perfettini2004postseismic}. It is in contrast with rupture, the stress-releasing motion, which is an unstable sliding that continues to accelerate, where the velocity accelerates as an exponential function of slip, or approximately of time, growing explosively near the velocity peak ~\cite{dieterich1986model}. Those collective pieces of evidence raise the hypothesis that the primary slip mode of the 2010 Bungo Channel SSE is postseismic creep rather than coseismic rupture, a hypothesis we examine below.

\subsection*{Trend fitting of stacked time series}

Stacked data confirm that the postseismicwise group (Fig.~\ref{fig:1}b blue) also detect a run-up signal prior to recording the blow-up signal as the co-\&post-seismicwise group does (Fig.~\ref{fig:1}b purple). 
The only difference from the postseismicwise group from the co-\&post-seismicwise group is the level of the preceding run-up signal. 
Those groups are in contrast to the preseismicwise group that tracks a gradually increasing signal throughout the analysis period even in stacked data. 
To objectively separate the multiple signals observed, we perform a functional fit to the stacked data.

We determine the onsets of the log trend and the preceding small run-up trend 
by applying functional fitting to the stacked postseismicwise time series and the stacked coseismicwise time series, which clearly detect run-up and blow-up trends. 
To confirm the nonlocal natures of these two trends, the functional fit is also applied to the stacked time series of all observation stations shown in Fig.~\ref{fig:1}a. 
The preseismicwise trend, which appears locally only on the west (Fig.~\ref{fig:1}a), will be discussed in the next subsection. The functional fitting employs least-square analysis with the following two empirical functions: one is a simple linear-plus-log piecewise function, where the onsets of coseismicwise and postseismicwise signals are well-defined, 
\begin{equation}
    F_1(t)=\left\{
\begin{array}{ll}
0 & (t<t_1)
\\
c_1(t-t_1)& (t_1<t<t_2)
\\ 
c_1(t_1-t_0)+c_2\ln [(t-t_2)/t_{\rm c}+1]& (t_2<t)
\end{array}
\right.
\end{equation}
and the other is a superposition of a logistic function and a log function, which expresses the mixed signals of the smooth exponential growth and the abruptly starting logarithmic rise,
\begin{equation}
    F_2(t)=c_1^\prime\left(1+Ke^{-t/t_*}\right)^{-1}+c_2^\prime \ln[R(t-t_2)/t_{\rm c}+1],
\end{equation}
where $t_1$ and $t_2$ represent the onset times of the coseismicwise and postseismicwise trends, respectively, and $t_{\rm c}$ denotes the cutoff time of the postseismicwise logarithmic evolution; ($c_1,c_2,c_1^\prime, c_2^\prime$) denote prefactors of respective trends, $R(\cdot)$ denotes the ramp function, and $(K,t_*)$ are parameters of the logistic function. 
The logistic function is a regular function and models an exponential growth near $t\sim t_*\ln K$, and thus eq.~(2) expresses the superposition of exponential and log trends. 
To account for the potential contamination of the postseismicwise signal by the coseismicwise signal, 
we define the end of the coseismicwise signal ($t_2^{\prime}$) 
as the time when the magnitudes of the logistic function and log function become equal in eq.~(2), after which the exponential trend becomes negligible relative to the log trend. 

\begin{figure}
	\centering
	\includegraphics[width=0.95\textwidth]{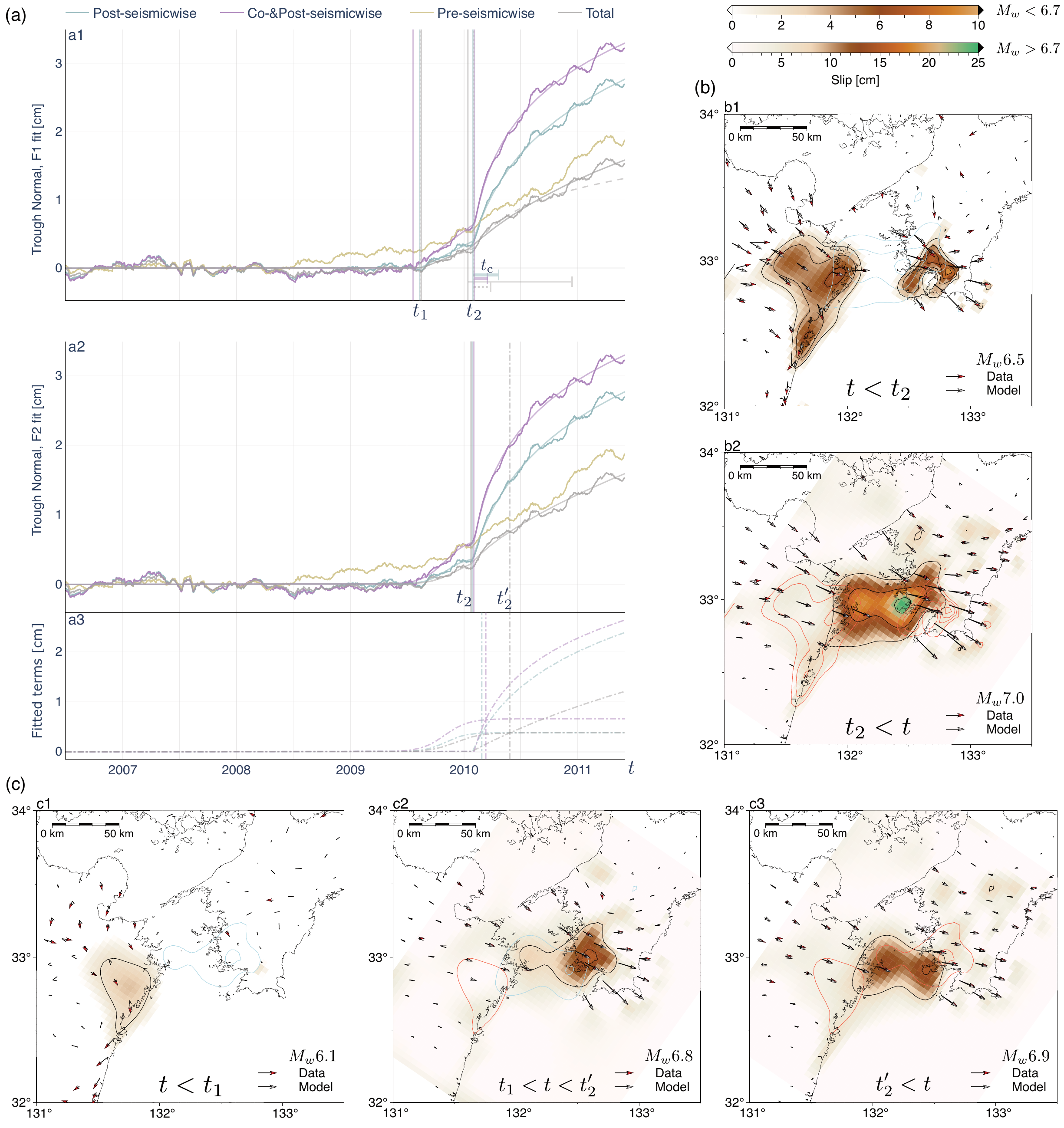}
  \caption{
  Time-domain decomposition and slip inversion of decomposed signals. 
  Time-domain decomposition is based on least-square functional fitting, and slip inversion evaluates the most probable estimate using the prior constraint of the Elastic Net. 
  Estimated slips are visualized by
  two different color scales, distinguished by whether the moment-magnitude estimate is below or above 6.7, assuming a rigidity of $30 \mathrm{~GPa}$; the contours of the preceding and subsequent events are drawn in blue and red, respectively. 
  (a) Partitioning of coseismicwise and postseismicwise time domains based on functional fitting (a1: eq. 1; a2: eq. 2). The colors of the lines follow Fig.~\ref{fig:1}b, and the fitted curves are shown in the corresponding light colors. 
  Vertical solid lines indicate the estimated onset times of the coseismicwise and postseismicwise signals ($t_1$ and $t_2$), and vertical dash-dots indicate the estimated termination time of the coseismicwise signal $t_2^\prime$ set by eq.~(2), which is the intersection of the sigmoidal growth and the logarithmic rise (a3, dash-dot curves). Horizontal lines in Fig.~\ref{fig:2}a1 represent the cutoff time $t_c$ of the log slip. 
  The dotted lines in Fig.~\ref{fig:2}a1 indicate the fitting of eq.~(1) for the data period before the end of October 2010, after which the eastern Shikoku SSE may contaminate the data. 
  (b) Inversion of logarithmic signals and preceding signals, separated at time $t_2$, February 2010. (c) Inversion of GNSS data separated at the onset time of the coseismicwise signal $t_1$, July 2009, and its termination time $t_2^\prime$, May 2010.
}
  \label{fig:2}
\end{figure}

The least-square solutions are shown in Fig.~\ref{fig:2}a, where the fitting curves of light colors match the stacked data of dark colors well. 
Vertical and horizontal straight lines indicate the associated characteristic times ($t_1$, $t_2$, $t_{\rm c}$, $t_2^\prime$), whose specific values are summarized in Supplementary Tables~S1--S3. 
In both the linear-plus-log fit (eq.~1, Fig.~\ref{fig:2}a1) and the exponential-plus-log fit (eq.~2, Fig.~\ref{fig:2}a2), every fitting curve detects the coseismicwise trend from mid-2009 and the postseismicwise trend from early 2010.

These fitting curves locate almost the same signal onsets within the accuracy range of the one-month low-pass used for data processing (Fig.~\ref{fig:2}a, vertical solid lines). 
The onset of the coseismicwise trend $t_1$ in mid-2009 is late July to mid-August for the linear-plus-log fit (eq.~1). 
The onset of the postseismicwise trend $t_2$ in early 2010 is mid-January to early February for the linear-plus-log fit (eq.~1) 
and late January to early February for the exponential-plus-log fit (eq.~2). 
The onset of the coseismicwise trend is estimated to be earlier when the coseismicwise signal is large, suggesting better detectability for a high SN ratio. 
For the logarithmic rise, which produces the strongest signal at the onset, the onset time is consistently estimated to be almost the same value, February 1st, regardless of the function and data used, except when the stacking of all stations is fitted with a linear-plus-log function.

The above functional fits show that coseismicwise and postseismicwise signals well separate. 
The logarithm and the east exponential are consistent with the activation of slip beneath the Bungo Channel from around February 2010 and the event near the north of Cape Ashizuri preceding the log slip by about half a year, respectively. 

An extremely large $t_c$ is obtained only for the total signal (Fig.~\ref{fig:2}a1, horizontal solid line, purple), and this could be attributed to another SSE that occurred around central Shikoku from late October 2010~\cite{takagi2019along}. When refitting is performed by using the truncated time window cutting the last half a year, which is on or before the November 2010, $t_c$ is found to be about two months, consistent with the other estimates of $t_c$ (Fig.~\ref{fig:2}a1, horizontal dotted line, purple; also see Supplementary Table~S2). 
It is considered that a similar log trend is detected spatially uniformly. 
The termination time of the coseismicwise signal, $t_2^\prime$, is determined to be July 2010 at most (Figs.~\ref{fig:2}a2 and \ref{fig:2}a3, vertical dash dots). 
Here, $t_c$ (Fig.~\ref{fig:2}a1, horizontal lines) roughly corresponds to the difference between $t_2$ and the maximum $t_2^\prime$, meaning that the exponential slip has mostly finished after $t-t_2>t_{\rm c}$, where the signal decelerates as the cutoff of the log becomes negligible.

\subsection*{Sparse inverse modeling}
The goodness of the functional fitting leads to the conclusion that the major signal in the Bungo Channel is a logarithmic slip with a cutoff time of approximately two months.
Spatial distributions of the station groups further suggest that the log slip in the channel follows two distinct exponential slips: a west gradual slip lasting about two years and an east slip lasting about half a year.
The gradual preseismicwise signal in the west, the third trend of the GNSS data, is not temporally isolated from the other two signals, which suggests a slip episode inconsistent with the previous interpretation of the 2010 SSE as a slow-slip migration from the east to the channel. 
We therefore constrain the spatiotemporal relationships of those slip episodes by performing slip inversions that account for the characteristic time domains of the signals.

The estimates of $t_1$, $t_2$, and $t_c$ consistently show identical values, suggesting that they represent characteristic moments and a duration of the actual source process, rather than merely being artificial parameters of functional fits. We thus partition the time range by $t_1$, $t_2$, and $t_2^\prime\sim t_2+t_{\rm c}$ to invert the horizontal displacement data. 
Here, $t_1$ and $t_2$ refer to the least-square fits of the co-\&post-seismicwise stacked data where those signal onsets become the most visible, corresponding to July 2009 and February 2010, respectively. The highly variable $t_2^\prime$ is set to 
90 days, around 1.5 times $t_c$, following $t_2$, 
which fits the maximum $t_2^\prime$ estimate in month units, May 2010. 
Because the spatial distribution of slip is smeared out in conventional inverse analysis, which constrains the weighted L2 norms of model parameters, we employ sparse modeling that incorporates an L1 constraint~\cite{zou2005regularization} to delineate the spatial relationships of slip distributions at different windows.
Inversion details are summarized in Methods.

Fig.~\ref{fig:2}b shows the most probable estimate of the slip distribution regarding the cumulative slip before and after the onset time of the logarithmic slip $t_2$. The cumulative slip after $t_2$ ($t_2 < t$, $M_w = 7.0$), which has been recognized as the main part of the Bungo Channel SSE, peaks in the channel, confirming that the logarithmic signal originates from the SSE right beneath the Bungo Channel. The slip distribution preceding the log ($t < t_2$, $M_w = 6.5$) is separated into the regions east and west of the channel, corresponding to the two temporal trends of preseismicwise and coseismicwise signals observed in the east and west stations preceding the log (Fig.~\ref{fig:1}a). 

Fig.~\ref{fig:2}c presents the most probable estimates when partitioned by $t_1$, the linear trend onset of the coseisimicwise signal, and $t_2^\prime$, the termination time of the the coseisimicwise exponential signal. The signal preceding the linear trend is predominantly fault slip west of the channel ($M_w=6.1$, $t<t_1$). 
After the onset of the linear trend $t_1$, the eastern slip becomes more active, surpassing the western slip ($t_1<t<t_2^\prime$).

As noticed by comparing the slips for $t<t_1$ and for $t<t_2$, the western side is still moving when the eastern side starts activity ($t_1<t<t_2$). However, as later quantified in Fig.~\ref{fig:3}b, the western slip is negligible compared to the active eastern slip ($M_w6.8$) during $t_1<t<t_2^{\prime}$ including the early period of the log slip. 
Crucially, there is a spatial discontinuity of the main slip zones between the west-dominant period $t<t_1$ and 
the east-dominant period $t_1<t<t_2^{\prime}$. 
The estimate reveals that spatially close but separated exponential growths successively activated, followed by a logarithmic slip ($M_w6.9\sim7.0$) beneath the channel.

The remaining decompositions of $t_1<t<t_2$ and $t_2<t<t_2^\prime$ are supplemented in Fig.~S1. The slip of the former case $t_1<t<t_2$ is less constrained as it is the inference of relatively small slips susceptible to data noise. In the text, the slips during those intervals are calculated by synthesizing relatively robust estimates shown in Fig.~\ref{fig:2} (Fig.~\ref{fig:3}b). 

\section*{Discussion}

The logarithmic signal of the 2010 Bungo Channel SSE lets us discern that a slow earthquake also entails a postseismic phase, motivating this study to test whether the observed signal truly derives from afterslip. 
Estimated snapshots of the fault slip for characteristic periods indicate that the Bungo Channel SSE thus-far recognized as a single event includes three episodes: the Bungo area first hosts (A) a longer-term event of exponential growth lasting nearly two years west, adjacent to the channel, which is followed by (B) a relatively short exponential growth of about half a year east, adjacent to the channel, finally leading to (C) a logarithmic slip beneath the channel.
The fault slips of respective episodes are close in space and time, but they differ in onset times and source regions.
Even the episode in the west, poorly separated from the other episodes in the observed signals, clearly separates in the fault-slip estimate. 
By performing sparse modeling in the individual time periods that produce characteristic trends of signals, we were able to separate multiple slip episodes that were obscure in conventional analyses.

\subsection*{Correspondence to repeating slow-slip events and synchronized tremor bursts}

\begin{figure}
	\centering
	\includegraphics[width=1.0\textwidth]{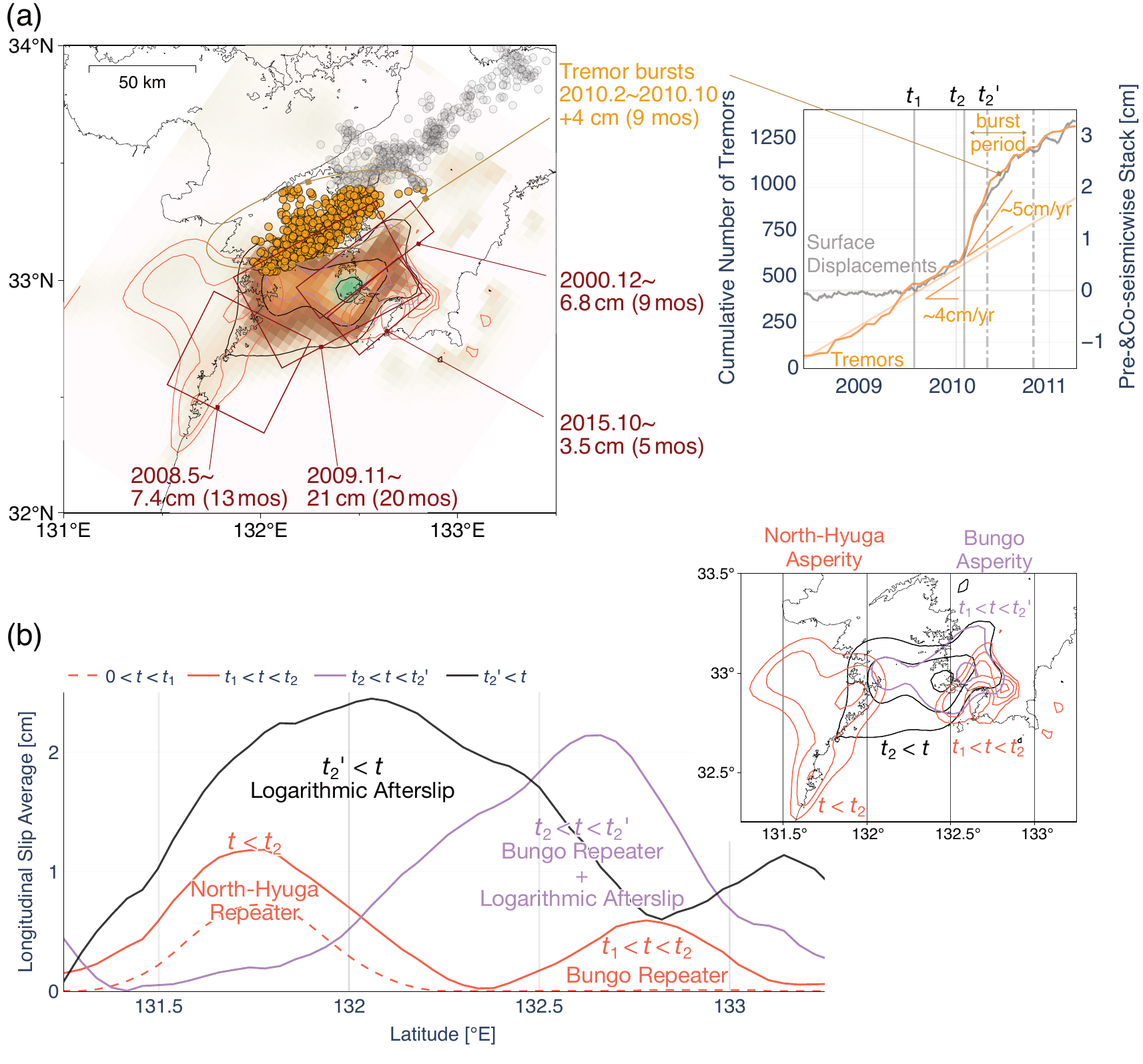}
  \caption{
  Comparison between the inferred 2010 SSE and slow-earthquake records, and a Bungo slow-earthquake synthesis. 
  (a) Estimated slip zones of the 2010 event and documented slow earthquakes. 
  The estimated slip follows the result shown in Fig.~\ref{fig:2}b2, where the red contour outlines the exp slips that precede the log slip ($t<t_2$), while the color map represents the log slip ($t>t_2$). 
  The slow-earthquake record consists of the estimated SSE areas [red rectangles~\cite{takagi2019along}] and the tremor events (gray circles) that occurred between February and October~\cite{annoura2016total}. 
  For the synchronized events south of 33.4$^{\circ}$N (orange circles), the slip distance due to burst events exceeding the offset rate is calculated from the radiated energy. 
  The number of tremors and its offset rate (straight line) are shown alongside the stacked pre-\&co-seismicwise signals (gray, from Fig.~\ref{fig:1}b purple). 
  (b) Interpretations of the estimated episodes, fitted to the Bungo slow-earthquake cycle. Longitudinal slip averages for latitudinal bins of 10 km are calculated from slip inversions. 
  Exponential growths and the primary parts of logarithmic slips are respectively interpreted as coseismic ruptures and postseismic creeps, namely afterslips. An early part of the log trend is interpreted as a compound of rupture and creep, considering repeater locations and source mechanics. Slip contours obtained in Fig.~\ref{fig:2}b for $t<t_2$ and $t_2<t$ and Fig.~\ref{fig:2}c2 for $t_1<t<t_2^\prime$ are drawn in the colors of the average slips for the time windows of the same ends. 
}
  \label{fig:3}
\end{figure}

Figure~\ref{fig:3}a is a comparison of our estimate with the record of the slow-earthquake activity around the Bungo Channel~\cite{annoura2016total,takagi2019along}. 
The western episode ($t<t_2$) took place at the same time and in the same location as the Northern Hyuga-nada SSE, and is therefore identified as the Northern Hyuga-nada SSE that is usually detrended in analyzing the Bungo Channel SSE.
The eastern episode occurring north of Cape Ashizuri is consistent with the source region of small events east the Bungo Channel estimated in 2000, 2014, and 2015, suggesting that this location activated also in 2010. 
The slip area to which the slip migrated northward from the vicinity of Cape Ashizuri during $t_2<t<t_2^\prime$ matches the rupture area from December 2000. The terminal signal of the exponential growth, mixed into the early stage of the log slip, slightly captures the slip prominent in this region.

A characteristic feature of the Bungo Channel SSE is the tremor synchronization responding to the magnitude of the slow slip~\cite{hirose2010slow,hirose2023long}, which was also the case for the 2010 event~\cite{ozawa2013spatial,nakata2017discontinuous}. 
The tremor activity south of 33.4$^{\circ}$N from February to October 2010, synchronized with the SSE presented here, is overlaid in Fig.~\ref{fig:3}a along with its temporal evolution. 
Although previously under-recognized, the tremor activity is not activated during the western and eastern exponential slips, but selectively responds and bursts during the log slip beneath the channel. 

The areal radiated-energy density of deep tremors for the associated spatiotemporal range is about $10^6$ J/km$^2$~\cite{annoura2016total}, which corresponds to a slip of about 8 cm for a tremor scaled energy of $4\times10^{-10}$~\cite{ide2008bridging} and an off-fault rigidity of 30 GPa. 
The tremor event rate is about twice, or approximately 9/4 times, the stationary rate during the burst period (Fig.~\ref{fig:3}a). Assuming 9/4 times the stationary rate is approximately an 8 cm slip over the 10-month burst, the slip rate is estimated to have accelerated during the burst from a steady-state rate of approximately 4 cm/year to about 9 cm/year; the addition is near 5 cm/year on average. 
The slip in the tremor burst area in the present slip inversion, which subtracts the inter-event rate, correlates well with this 4 cm slip calculated from the tremor burst. 
It leads to a conclusion that the slip in the deeper portion of the 2010 SSE source region is largely attributed to the seismic moment of the tremor, implying that tremor patches relatively densely fill the deeper portion of the Bungo Channel.

\subsection*{A synthesized interpretation of the Bungo slow-earthquake cycle}
Synthesizing the above findings, namely the temporal evolution of GNSS data, the spatial relationship between slip episodes, tremor synchronization selectively responding to the log slip, and the positional consistency of slip episodes with the SSE record, we interpret the 2010 Bungo Channel SSE as shown in Fig.~\ref{fig:3}b. 
Longitudinal averages of slip distributions 
are calculated for the characteristic periods $t<t_1$, $t_1<t<t_2$, $t_2<t<t_2^\prime$, and $t_2^\prime<t$ from additions and subtractions of the slip estimates shown in Figs.~\ref{fig:2}b and \ref{fig:2}c; the same qualitative behaviors are obtained from the slip estimates for respective time windows (Fig. S2).
Based on the SSE record, we read the observed exponential growths as repeating ruptures capable of occurring spontaneously. Specifically, the apparently-spontaneous western Northern Hyuga event ($t<t_2$) and the apparently-triggered eastern Cape Ashizuri episode ($t_1<t<t_2$) correspond to coseismic ruptures. 
The subsequent log slip ($t_2<t$) can be mostly interpreted as afterslip triggered by the preceding slow-rupture doublet, and the synchronized tremor burst as aftershocks. 
However, the slip north of Cape Ashizuri that occurs during the early stage of the log slip ($t_2<t<t_2^\prime\sim t_2+t_{\rm c}$) is more consistent with the SSE record if it is read as including an extension of the Cape Ashizuri episode. 
Therefore, we interpret the log slip not as the pure creep initially thought, but as a compound of postseismic creep and coseismic rupture in its early stage ($t_2<t<t_2^\prime$). 
We suppose that creep and rupture do not clearly separate in that transition period because the slip rate of an SSE is not as fast as that of an ordinary earthquake. 
If the coseismic slip signal is mixed within the early afterslip signal,
it is natural that the afterslip in this event appears faster and is accompanied by a larger slip distance than the rupture doublet.
Even in a mechanical viewpoint, afterslip is more active in the source region during the early stage of the postseismic phase~\cite{kato2004interaction}, so it is physically natural that the source region of afterslip initially coincides with the main rupture zone. 
In the present estimate, as expected, the main area of the log slip shifts from the center of the Bungo repeater to the surrounding area around the boundary $t_2^{\prime}$, appearing as if the afterslip diffuses outward from the Bungo repeater as the coseismic source area (Figs. \ref{fig:2}b2, \ref{fig:2}c3, and \ref{fig:3}b).

The kinematic observable is only the fault motion, and thus it is not possible to conclude solely from the present slip inversion that the log slip is afterslip.
However, the prevailing interpretation of the 2010 Bungo SSE as rupture migration is likely an artifact resulting from the exclusion of the unmodeled Northern Hyuga-nada SSE. 
Even in existing inverse analyses of the 2010 Bungo SSE, a remnant of the Northern Hyuga SSE was faintly visible but was located at the edge of the model fault~\cite{ozawa2013spatial} and was smoothed, meaning that the 2010 SSE could only be expressed as migration from the east toward the Bungo Channel~\cite{hirose2010source,yoshioka2015spatiotemporal}. 
Although there are reports that the SSE slip rate accelerates when the SSE rupture fronts coalesce~\cite{bletery2020slip}, the eastern event activates after the cessation of the western event in our analysis, and there is not a strong motivation to interpret the slip behavior as a coalescence in the present kinematic modeling. 
The results of this study do not reject the null hypothesis that the log signal does not originate from afterslip, but they are supportive of the above interpretation that postseismic creep is included in the SSE and was observed this time. 

Even in previously proposed physics-based models, the areas of conditionally unstable friction capable of generating SSEs are estimated to prevail north of Cape Ashizuri, while frictionally stable zones are dominant near the channel where large-scale slip occurred~\cite{fukushima2025spatio}. 
Although the SSE estimate appears migratory even in those analyses because the western episode of the Northern Hyuga SSE is an unmodeled slip outside their model faults, the estimate of frictional parameters, where the primary asperity is on the Cape-Ashizuri side rather than the Bungo Channel, directly supports our interpretation that postseismic creep is dominant beneath the channel.

Afterslips exceeding the coseismic magnitudes, including that for an adjacent rupture multiplet, have also been reported for ordinary earthquakes~\cite{heki1997silent,ozawa2012preceding,churchill2022afterslip}. 
The moment release of the exponential slips of Northern Hyuga and North Cape Ashizuri detected here corresponds to those of 2008.5 and 2000.12 SSEs [Mw 6.5 + Mw 6.3~\cite{takagi2019along}], while that of the log slip beneath the Channel is Mw 6.9. 
Their postseismic-coseismic moment ratio is about 2.7, which is consistent with the moment ratio 2.5 between the afterslip and the coseismic slips of the Mw 7 class earthquake quartet that occurred in the Japan trench in 2003--2010~\cite{ozawa2012preceding}.

During the maximum acceleration of coseismic slip, the slip rate is expected to explode, governed by the inverse of the time it takes to reach the maximal speed, the inverse Omori law~\cite{dieterich1986model}. Therefore, even the observed acceleration during the exponential-logarithmic transition $t_2<t\lesssim t_2+t_{\rm c}$ where the log exceeds the top speed of the exponential growth does not negate the present interpretation. It just implies the possibility of coseismic slips faster than exponentials within that transient.

On record~\cite{ozawa2024time}, the Bungo Channel has experienced slow slips of different magnitudes, Mw$\sim$7 and Mw$\sim$6.3.
The present interpretation of the primary source of the 2010 event, active afterslip following the eastern Channel rupture, enhanced by the precedence of the Northern Hyuga SSE, provides one mechanical explanation for the event magnification, a hypercycle, of the Bungo slow earthquake cycle. 
Since the Bungo Channel is located at the westernmost part of the Nankai Trough, such cyclic slow slips with variable magnitudes are analogous to the initial process of a megathrust earthquake, where the edge of the giant locked zone recursively erodes in variable magnitudes, and the surrounding area slowly creeps~\cite{ohnaka1992earthquake,liu2007spontaneous,noda2013large}.

\subsection*{Model implications derived from postseismicity hypothesis}
Postseismic creep emerges inevitably when simulating spontaneous SSEs, yet it has not been supported by observations.
In order to align with observational reports, the physics-based reproduction of SSEs has reluctantly relied on artificial initial conditions to preclude postseismic creep, such as zero-coupling conditions at the immediate vicinity of the locked zone~\cite{kano2024data}, or simply neglected such a presence. 
Explicitly stating the mechanical ubiquity of the postseismicity that extends to slow earthquakes is relevant to discussing the epistemic error in physics-based slow-earthquake modeling, that is, the candidate rupture processes implicitly excluded from the model space. 
A logarithmic slip trend indicates that the source process has entered a deceleration phase, allowing log SSEs to be distinguished from rupture growths that could accelerate exponentially toward disruptive failure. 
The recognition that there are qualitatively different SSEs, characterized by logarithmic deceleration and exponential acceleration, provides a useful viewpoint for discussing the reproducibility~\cite{rubin2008episodic,kano2024data} and forecastability ~\cite{liu2009slow,truttmann2024slow,kaveh2025spatiotemporal} of slow earthquakes, integrating them with precursory nucleations of large earthquakes.

During slow slip, which is merely a slow fault motion in kinematics, it is a mechanical expectation that both slow rupture and triggered creep occur if slow rupture exists, although to our knowledge, it has been under recognized thus far. 
The underlying mechanism is so robust: the surrounding creeping zone accelerates once the locked zone ruptures, and gradually decelerating creep remains in a postseismic phase where the rupture ceases.
The observation of this robust mechanics has likely been prevented by a known shortcoming of slip inversion: afterslip tends to artificially shrink around the coseismic rupture zone~\cite{sherrill2021new}, which makes the decomposition of coseismic rupture and postseismic creep even more challenging in SSEs where a velocity scale separation of rupture and creep is improbable.
A rupture doublet enhances afterslip and shifts the central area of the creep away from the rupture-zone fringe toward the doublet's center, thereby facilitating the observational detection of afterslip. 
We deem that the Bungo Channel doublet thus generated postseismic creep at a non-negligible level. 
The 2010 Bungo Channel SSE prompts the recognition that it may be physically unnatural to dismiss the presence of the two different phases during the course of a slow earthquake.

\subsection*{Concluding remarks}
Once a locked zone ruptures, the surrounding creeping zone accelerates in tow and slowly decelerates in a postseismic phase while the rupture zone rapidly relocks.
Since slow earthquakes are faulting just like ordinary earthquakes, a postseismic phase of a slow earthquake should physically exist.
We have addressed this postseismic phase of slow earthquakes based on a reanalysis of the 2010 Bungo Channel long-term SSE. 
The observed crustal deformation is most rapid at the onset of the SSE, which defies the standard picture that draws the SSE as a rupture event similar to an ordinary earthquake except for contrasting slip rates.
The stacked data indicate that the primary signal is preceded by two minor signals of exponential growth within different durations. 
These preceding exponential signals bracket the channel roughly at the same depth, and the primary signal emanating from the channel follows the two logarithmically. 
By performing sparse modeling of GNSS data partitioned by the characteristic moments of those signals, we were able to detect a slow-slip doublet that grew exponentially on the east and west, preceding the log slip beneath the channel. 
The western SSE episode is the Northern Hyuga-nada SSE, usually unmodeled in analyses of the Bungo Channel SSE, and the eastern SSE episode matches the source region of a small Bungo Channel repeater. 
The logarithmic SSE proceeding beneath the channel, particularly its latter part, can be interpreted as postseismic slip triggered by this slow-rupture doublet, and the tremor burst synchronized with the rate of the log slip as aftershocks.
The sequence of slip episodes constructing the Bungo slow-earthquake cycle is understandable as a rupture doublet and triggered afterslip and aftershocks, analogous to the rupture multiplet and enhanced afterslip observed for ordinary earthquakes. 
While the actual source process of the Bungo slow-earthquake cycle remains open to debate, the postseismic phase in seismic cycles itself is mechanically expected, and the 2010 Bungo Channel SSE provides an important opportunity to realize that such a mechanical perspective has been missing in the literature thus far.
The estimated source process of the Bungo Channel SSE, which differs from a simple slow rupture, suggests that the strain release mode in the westernmost region of the Nankai subduction zone is a slow-rupture interplay followed by a slow postseismic creep. 
When such an aseismic slip erodes the outermost layer of the megathrust seismogenic zone, the action relevant to forecasting the evolution of slow slips is differentiating the coseismic and postseismic slips concealed underneath slow events. We consider the logarithm signal as the first clue to that distinction.

\section*{Methods}
In the inverse analysis, 
the model employs a rectangular fault, the center of which is positioned beneath the center of the Bungo Channel 32.9$^\circ$N and 132.3$^\circ$E at a depth of 25 km. The fault is oriented with a convergence direction of N57.7$^\circ$W and a dip angle of 15 degrees, following previous data-assimilation studies~\cite{kano2024data}. The model fault spans 240 km along strike and 200 km along dip, subdivided into 4 km square subfaults. We assume a uniform isotropic elastic half-space~\cite{okada1985surface} with the surface set at the sea level.
The inverse analysis is conducted based on a Bayesian sparse model using the Elastic Net~\cite{zou2005regularization}, which imposes a prior constraint defined as a linear combination of the L1 and L2 norms of model parameters. 
Here, the model parameters represent the slip distances on the subfaults. Weight coefficients of the Elastic Net, which are additional parameters, are estimated simultaneously via 10-fold cross-validation based on the $R^2$ score. 
The most probable estimate is searched as the optimal by using the coordinate descent.
The inverted data are horizontal components of cumulative displacements integrated over $t>0$, which is set at 2008-04-10, 100 days after 2008-01-01. 
Cumulative displacements are partitioned by the characteristic times $t=t_1$, $t_2$, and $t_2^\prime$, set at 2009-07-21 (the $t_1$ estimate for the co-\&post-seismicwise data), 2010-02-02 (the mean estimate of $t_2$ for the co-\&post-seismicwise data), and 2010-05-03 (90 days after $t_2$), respectively. 
In evaluating the likelihood, the data components are assumed to be statistically independent Gaussian random variables of an identical variance. 


\begin{thebibliography}{10}
\expandafter\ifx\csname url\endcsname\relax
  \def\url#1{\texttt{#1}}\fi
\expandafter\ifx\csname urlprefix\endcsname\relax\def\urlprefix{URL }\fi
\providecommand{\bibinfo}[2]{#2}
\providecommand{\eprint}[2][]{\url{#2}}

\bibitem{hirose1999slow}
\bibinfo{author}{Hirose, H.}, \bibinfo{author}{Hirahara, K.},
  \bibinfo{author}{Kimata, F.}, \bibinfo{author}{Fujii, N.} \&
  \bibinfo{author}{Miyazaki, S.}
\newblock \bibinfo{title}{{A slow thrust slip event following the two 1996
  Hyuganada earthquakes beneath the Bungo Channel, southwest Japan}}.
\newblock \emph{\bibinfo{journal}{Geophysical Research Letters}}
  \textbf{\bibinfo{volume}{26}}, \bibinfo{pages}{3237--3240}
  (\bibinfo{year}{1999}).

\bibitem{obara2002nonvolcanic}
\bibinfo{author}{Obara, K.}
\newblock \bibinfo{title}{{Nonvolcanic deep tremor associated with subduction
  in southwest Japan}}.
\newblock \emph{\bibinfo{journal}{Science}} \textbf{\bibinfo{volume}{296}},
  \bibinfo{pages}{1679--1681} (\bibinfo{year}{2002}).

\bibitem{miller2002periodic}
\bibinfo{author}{Miller, M.~M.}, \bibinfo{author}{Melbourne, T.},
  \bibinfo{author}{Johnson, D.~J.} \& \bibinfo{author}{Sumner, W.~Q.}
\newblock \bibinfo{title}{{Periodic slow earthquakes from the Cascadia
  subduction zone}}.
\newblock \emph{\bibinfo{journal}{Science}} \textbf{\bibinfo{volume}{295}},
  \bibinfo{pages}{2423--2423} (\bibinfo{year}{2002}).

\bibitem{rogers2003episodic}
\bibinfo{author}{Rogers, G.} \& \bibinfo{author}{Dragert, H.}
\newblock \bibinfo{title}{{Episodic tremor and slip on the Cascadia subduction
  zone: The chatter of silent slip}}.
\newblock \emph{\bibinfo{journal}{Science}} \textbf{\bibinfo{volume}{300}},
  \bibinfo{pages}{1942--1943} (\bibinfo{year}{2003}).

\bibitem{yagi2003partitioning}
\bibinfo{author}{Yagi, Y.} \& \bibinfo{author}{Kikuchi, M.}
\newblock \bibinfo{title}{{Partitioning between seismogenic and aseismic slip
  as highlighted from slow slip events in Hyuga-nada, Japan}}.
\newblock \emph{\bibinfo{journal}{Geophysical research letters}}
  \textbf{\bibinfo{volume}{30}} (\bibinfo{year}{2003}).

\bibitem{kato2012propagation}
\bibinfo{author}{Kato, A.} \emph{et~al.}
\newblock \bibinfo{title}{{Propagation of slow slip leading up to the 2011 M w
  9.0 Tohoku-Oki earthquake}}.
\newblock \emph{\bibinfo{journal}{Science}} \textbf{\bibinfo{volume}{335}},
  \bibinfo{pages}{705--708} (\bibinfo{year}{2012}).

\bibitem{kanamori1974focal}
\bibinfo{author}{Kanamori, H.} \& \bibinfo{author}{Cipar, J.~J.}
\newblock \bibinfo{title}{{Focal process of the great Chilean earthquake May
  22, 1960}}.
\newblock \emph{\bibinfo{journal}{Physics of the Earth and Planetary
  Interiors}} \textbf{\bibinfo{volume}{9}}, \bibinfo{pages}{128--136}
  (\bibinfo{year}{1974}).

\bibitem{ito2013episodic}
\bibinfo{author}{Ito, Y.} \emph{et~al.}
\newblock \bibinfo{title}{{Episodic slow slip events in the Japan subduction
  zone before the 2011 Tohoku-Oki earthquake}}.
\newblock \emph{\bibinfo{journal}{Tectonophysics}}
  \textbf{\bibinfo{volume}{600}}, \bibinfo{pages}{14--26}
  (\bibinfo{year}{2013}).

\bibitem{bletery2023precursory}
\bibinfo{author}{Bletery, Q.} \& \bibinfo{author}{Nocquet, J.-M.}
\newblock \bibinfo{title}{{The precursory phase of large earthquakes}}.
\newblock \emph{\bibinfo{journal}{Science}} \textbf{\bibinfo{volume}{381}},
  \bibinfo{pages}{297--301} (\bibinfo{year}{2023}).

\bibitem{kawasaki2001space}
\bibinfo{author}{Kawasaki, I.}, \bibinfo{author}{Asai, Y.} \&
  \bibinfo{author}{Tamura, Y.}
\newblock \bibinfo{title}{{Space--time distribution of interplate moment
  release including slow earthquakes and the seismo-geodetic coupling in the
  Sanriku-oki region along the Japan trench}}.
\newblock \emph{\bibinfo{journal}{Tectonophysics}}
  \textbf{\bibinfo{volume}{330}}, \bibinfo{pages}{267--283}
  (\bibinfo{year}{2001}).

\bibitem{ozawa2002detection}
\bibinfo{author}{Ozawa, S.} \emph{et~al.}
\newblock \bibinfo{title}{{Detection and monitoring of ongoing aseismic slip in
  the Tokai region, central Japan}}.
\newblock \emph{\bibinfo{journal}{Science}} \textbf{\bibinfo{volume}{298}},
  \bibinfo{pages}{1009--1012} (\bibinfo{year}{2002}).

\bibitem{shelly2006low}
\bibinfo{author}{Shelly, D.~R.}, \bibinfo{author}{Beroza, G.~C.},
  \bibinfo{author}{Ide, S.} \& \bibinfo{author}{Nakamula, S.}
\newblock \bibinfo{title}{{Low-frequency earthquakes in Shikoku, Japan, and
  their relationship to episodic tremor and slip}}.
\newblock \emph{\bibinfo{journal}{Nature}} \textbf{\bibinfo{volume}{442}},
  \bibinfo{pages}{188--191} (\bibinfo{year}{2006}).

\bibitem{wech2007cascadia}
\bibinfo{author}{Wech, A.~G.} \& \bibinfo{author}{Creager, K.~C.}
\newblock \bibinfo{title}{{Cascadia tremor polarization evidence for plate
  interface slip}}.
\newblock \emph{\bibinfo{journal}{Geophysical Research Letters}}
  \textbf{\bibinfo{volume}{34}} (\bibinfo{year}{2007}).

\bibitem{ide2007mechanism}
\bibinfo{author}{Ide, S.}, \bibinfo{author}{Shelly, D.~R.} \&
  \bibinfo{author}{Beroza, G.~C.}
\newblock \bibinfo{title}{{Mechanism of deep low frequency earthquakes: Further
  evidence that deep non-volcanic tremor is generated by shear slip on the
  plate interface}}.
\newblock \emph{\bibinfo{journal}{Geophysical Research Letters}}
  \textbf{\bibinfo{volume}{34}} (\bibinfo{year}{2007}).

\bibitem{nishikawa2019slow}
\bibinfo{author}{Nishikawa, T.} \emph{et~al.}
\newblock \bibinfo{title}{{The slow earthquake spectrum in the Japan Trench
  illuminated by the S-net seafloor observatories}}.
\newblock \emph{\bibinfo{journal}{Science}} \textbf{\bibinfo{volume}{365}},
  \bibinfo{pages}{808--813} (\bibinfo{year}{2019}).

\bibitem{kostrov1988principles}
\bibinfo{author}{Kostrov, B.~V.} \& \bibinfo{author}{Das, S.}
\newblock \emph{\bibinfo{title}{Principles of earthquake source mechanics}}
  (\bibinfo{publisher}{Cambridge University Press}, \bibinfo{year}{1988}).

\bibitem{avouac2015geodetic}
\bibinfo{author}{Avouac, J.-P.}
\newblock \bibinfo{title}{From geodetic imaging of seismic and aseismic fault
  slip to dynamic modeling of the seismic cycle}.
\newblock \emph{\bibinfo{journal}{Annual Review of Earth and Planetary
  Sciences}} \textbf{\bibinfo{volume}{43}}, \bibinfo{pages}{233--271}
  (\bibinfo{year}{2015}).

\bibitem{perfettini2004postseismic}
\bibinfo{author}{Perfettini, H.} \& \bibinfo{author}{Avouac, J.-P.}
\newblock \bibinfo{title}{{Postseismic relaxation driven by brittle creep: A
  possible mechanism to reconcile geodetic measurements and the decay rate of
  aftershocks, application to the Chi-Chi earthquake, Taiwan}}.
\newblock \emph{\bibinfo{journal}{Journal of Geophysical Research: Solid
  Earth}} \textbf{\bibinfo{volume}{109}} (\bibinfo{year}{2004}).

\bibitem{kato2004interaction}
\bibinfo{author}{Kato, N.}
\newblock \bibinfo{title}{{Interaction of slip on asperities: Numerical
  simulation of seismic cycles on a two-dimensional planar fault with
  nonuniform frictional property}}.
\newblock \emph{\bibinfo{journal}{Journal of Geophysical Research: Solid
  Earth}} \textbf{\bibinfo{volume}{109}} (\bibinfo{year}{2004}).

\bibitem{kato2007expansion}
\bibinfo{author}{Kato, N.}
\newblock \bibinfo{title}{Expansion of aftershock areas caused by propagating
  post-seismic sliding}.
\newblock \emph{\bibinfo{journal}{Geophysical Journal International}}
  \textbf{\bibinfo{volume}{168}}, \bibinfo{pages}{797--808}
  (\bibinfo{year}{2007}).

\bibitem{helmstetter2009afterslip}
\bibinfo{author}{Helmstetter, A.} \& \bibinfo{author}{Shaw, B.~E.}
\newblock \bibinfo{title}{Afterslip and aftershocks in the rate-and-state
  friction law}.
\newblock \emph{\bibinfo{journal}{Journal of geophysical Research: Solid
  earth}} \textbf{\bibinfo{volume}{114}} (\bibinfo{year}{2009}).

\bibitem{kato20162016}
\bibinfo{author}{Kato, A.}, \bibinfo{author}{Nakamura, K.} \&
  \bibinfo{author}{Hiyama, Y.}
\newblock \bibinfo{title}{The 2016 {K}umamoto earthquake sequence}.
\newblock \emph{\bibinfo{journal}{Proceedings of the Japan Academy, Series B}}
  \textbf{\bibinfo{volume}{92}}, \bibinfo{pages}{358--371}
  (\bibinfo{year}{2016}).

\bibitem{cornet1997seismic}
\bibinfo{author}{Cornet, F.}, \bibinfo{author}{Helm, J.},
  \bibinfo{author}{Poitrenaud, H.} \& \bibinfo{author}{Etchecopar, A.}
\newblock \bibinfo{title}{Seismic and aseismic slips induced by large-scale
  fluid injections}.
\newblock \emph{\bibinfo{journal}{Pure and applied geophysics}}
  \textbf{\bibinfo{volume}{150}}, \bibinfo{pages}{563--583}
  (\bibinfo{year}{1997}).

\bibitem{peng2009remote}
\bibinfo{author}{Peng, Z.}, \bibinfo{author}{Vidale, J.~E.},
  \bibinfo{author}{Wech, A.~G.}, \bibinfo{author}{Nadeau, R.~M.} \&
  \bibinfo{author}{Creager, K.~C.}
\newblock \bibinfo{title}{{Remote triggering of tremor along the San Andreas
  Fault in central California}}.
\newblock \emph{\bibinfo{journal}{Journal of Geophysical Research: Solid
  Earth}} \textbf{\bibinfo{volume}{114}} (\bibinfo{year}{2009}).

\bibitem{materna2019dynamically}
\bibinfo{author}{Materna, K.}, \bibinfo{author}{Bartlow, N.},
  \bibinfo{author}{Wech, A.}, \bibinfo{author}{Williams, C.} \&
  \bibinfo{author}{B{\"u}rgmann, R.}
\newblock \bibinfo{title}{{Dynamically triggered changes of plate interface
  coupling in Southern Cascadia}}.
\newblock \emph{\bibinfo{journal}{Geophysical Research Letters}}
  \textbf{\bibinfo{volume}{46}}, \bibinfo{pages}{12890--12899}
  (\bibinfo{year}{2019}).

\bibitem{ozawa2003characteristic}
\bibinfo{author}{Ozawa, S.} \emph{et~al.}
\newblock \bibinfo{title}{{Characteristic silent earthquakes in the eastern
  part of the Boso peninsula, Central Japan}}.
\newblock \emph{\bibinfo{journal}{Geophysical Research Letters}}
  \textbf{\bibinfo{volume}{30}} (\bibinfo{year}{2003}).

\bibitem{bartlow2020long}
\bibinfo{author}{Bartlow, N.~M.}
\newblock \bibinfo{title}{{A long-term view of episodic tremor and slip in
  Cascadia}}.
\newblock \emph{\bibinfo{journal}{Geophysical Research Letters}}
  \textbf{\bibinfo{volume}{47}}, \bibinfo{pages}{e2019GL085303}
  (\bibinfo{year}{2020}).

\bibitem{wallace2020slow}
\bibinfo{author}{Wallace, L.~M.}
\newblock \bibinfo{title}{{Slow slip events in New Zealand}}.
\newblock \emph{\bibinfo{journal}{Annual Review of Earth and Planetary
  Sciences}} \textbf{\bibinfo{volume}{48}}, \bibinfo{pages}{175--203}
  (\bibinfo{year}{2020}).

\bibitem{bhattacharya2019fluid}
\bibinfo{author}{Bhattacharya, P.} \& \bibinfo{author}{Viesca, R.~C.}
\newblock \bibinfo{title}{Fluid-induced aseismic fault slip outpaces pore-fluid
  migration}.
\newblock \emph{\bibinfo{journal}{Science}} \textbf{\bibinfo{volume}{364}},
  \bibinfo{pages}{464--468} (\bibinfo{year}{2019}).

\bibitem{guglielmi2015seismicity}
\bibinfo{author}{Guglielmi, Y.}, \bibinfo{author}{Cappa, F.},
  \bibinfo{author}{Avouac, J.-P.}, \bibinfo{author}{Henry, P.} \&
  \bibinfo{author}{Elsworth, D.}
\newblock \bibinfo{title}{Seismicity triggered by fluid injection--induced
  aseismic slip}.
\newblock \emph{\bibinfo{journal}{Science}} \textbf{\bibinfo{volume}{348}},
  \bibinfo{pages}{1224--1226} (\bibinfo{year}{2015}).

\bibitem{rice1993spatio}
\bibinfo{author}{Rice, J.~R.}
\newblock \bibinfo{title}{Spatio-temporal complexity of slip on a fault}.
\newblock \emph{\bibinfo{journal}{Journal of Geophysical Research: Solid
  Earth}} \textbf{\bibinfo{volume}{98}}, \bibinfo{pages}{9885--9907}
  (\bibinfo{year}{1993}).

\bibitem{kano2024data}
\bibinfo{author}{Kano, M.}, \bibinfo{author}{Tanaka, Y.},
  \bibinfo{author}{Sato, D.}, \bibinfo{author}{Iinuma, T.} \&
  \bibinfo{author}{Hori, T.}
\newblock \bibinfo{title}{{Data assimilation for fault slip monitoring and
  short-term prediction of spatio-temporal evolution of slow slip events:
  application to the 2010 long-term slow slip event in the Bungo Channel,
  Japan}}.
\newblock \emph{\bibinfo{journal}{Earth, Planets and Space}}
  \textbf{\bibinfo{volume}{76}}, \bibinfo{pages}{57} (\bibinfo{year}{2024}).

\bibitem{fukushima2025spatio}
\bibinfo{author}{Fukushima, R.}, \bibinfo{author}{Kano, M.},
  \bibinfo{author}{Hirahara, K.} \& \bibinfo{author}{Ohtani, M.}
\newblock \bibinfo{title}{{Spatio-temporal evolution of the 2010 Bungo slow
  slip event revealed by Physics-Informed Neural Networks with rate and state
  friction}}.
\newblock In \emph{\bibinfo{booktitle}{JpGU 2025}} (\bibinfo{year}{2025}).

\bibitem{furumura2011revised}
\bibinfo{author}{Furumura, T.}, \bibinfo{author}{Imai, K.} \&
  \bibinfo{author}{Maeda, T.}
\newblock \bibinfo{title}{{A revised tsunami source model for the 1707 Hoei
  earthquake and simulation of tsunami inundation of Ryujin Lake, Kyushu,
  Japan}}.
\newblock \emph{\bibinfo{journal}{Journal of Geophysical Research: Solid
  Earth}} \textbf{\bibinfo{volume}{116}} (\bibinfo{year}{2011}).

\bibitem{hirose2023long}
\bibinfo{author}{Hirose, H.}, \bibinfo{author}{Matsushima, T.},
  \bibinfo{author}{Tabei, T.} \& \bibinfo{author}{Nishimura, T.}
\newblock \bibinfo{title}{{Long-term slow slip events with and without tremor
  activation in the Bungo Channel and Hyuganada, southwest Japan}}.
\newblock \emph{\bibinfo{journal}{Earth, Planets and Space}}
  \textbf{\bibinfo{volume}{75}}, \bibinfo{pages}{77} (\bibinfo{year}{2023}).

\bibitem{hirose2010source}
\bibinfo{author}{Hirose, H.}, \bibinfo{author}{Kimura, T.} \&
  \bibinfo{author}{Obara, K.}
\newblock \bibinfo{title}{{The source process of the 2009-2010 long-term slow
  slip event in the Bungo channel region based on Hi-net tilt and GEONET GPS
  data}}.
\newblock In \emph{\bibinfo{booktitle}{AGU Fall Meeting Abstracts}}, vol.
  \bibinfo{volume}{2010}, \bibinfo{pages}{S11C--08} (\bibinfo{year}{2010}).

\bibitem{ozawa2013spatial}
\bibinfo{author}{Ozawa, S.}, \bibinfo{author}{Yarai, H.},
  \bibinfo{author}{Imakiire, T.} \& \bibinfo{author}{Tobita, M.}
\newblock \bibinfo{title}{{Spatial and temporal evolution of the long-term slow
  slip in the Bungo Channel, Japan}}.
\newblock \emph{\bibinfo{journal}{Earth, Planets and Space}}
  \textbf{\bibinfo{volume}{65}}, \bibinfo{pages}{67--73}
  (\bibinfo{year}{2013}).

\bibitem{yoshioka2015spatiotemporal}
\bibinfo{author}{Yoshioka, S.}, \bibinfo{author}{Matsuoka, Y.} \&
  \bibinfo{author}{Ide, S.}
\newblock \bibinfo{title}{{Spatiotemporal slip distributions of three long-term
  slow slip events beneath the Bungo Channel, southwest Japan, inferred from
  inversion analyses of GPS data}}.
\newblock \emph{\bibinfo{journal}{Geophysical Journal International}}
  \textbf{\bibinfo{volume}{201}}, \bibinfo{pages}{1437--1455}
  (\bibinfo{year}{2015}).

\bibitem{dieterich1986model}
\bibinfo{author}{Dieterich, J.~H.}
\newblock \bibinfo{title}{A model for the nucleation of earthquake slip}.
\newblock \emph{\bibinfo{journal}{Earthquake source mechanics}}
  \textbf{\bibinfo{volume}{37}}, \bibinfo{pages}{37--47}
  (\bibinfo{year}{1986}).

\bibitem{takagi2019along}
\bibinfo{author}{Takagi, R.}, \bibinfo{author}{Uchida, N.} \&
  \bibinfo{author}{Obara, K.}
\newblock \bibinfo{title}{{Along-strike variation and migration of long-term
  slow slip events in the western Nankai subduction zone, Japan}}.
\newblock \emph{\bibinfo{journal}{Journal of Geophysical Research: Solid
  Earth}} \textbf{\bibinfo{volume}{124}}, \bibinfo{pages}{3853--3880}
  (\bibinfo{year}{2019}).

\bibitem{annoura2016total}
\bibinfo{author}{Annoura, S.}, \bibinfo{author}{Obara, K.} \&
  \bibinfo{author}{Maeda, T.}
\newblock \bibinfo{title}{{Total energy of deep low-frequency tremor in the
  Nankai subduction zone, southwest Japan}}.
\newblock \emph{\bibinfo{journal}{Geophysical Research Letters}}
  \textbf{\bibinfo{volume}{43}}, \bibinfo{pages}{2562--2567}
  (\bibinfo{year}{2016}).

\bibitem{hirose2010slow}
\bibinfo{author}{Hirose, H.} \emph{et~al.}
\newblock \bibinfo{title}{Slow earthquakes linked along dip in the {N}ankai
  subduction zone}.
\newblock \emph{\bibinfo{journal}{Science}} \textbf{\bibinfo{volume}{330}},
  \bibinfo{pages}{1502--1502} (\bibinfo{year}{2010}).

\bibitem{nakata2017discontinuous}
\bibinfo{author}{Nakata, R.} \emph{et~al.}
\newblock \bibinfo{title}{{Discontinuous boundaries of slow slip events beneath
  the Bungo Channel, southwest Japan}}.
\newblock \emph{\bibinfo{journal}{Scientific reports}}
  \textbf{\bibinfo{volume}{7}}, \bibinfo{pages}{6129} (\bibinfo{year}{2017}).

\bibitem{ide2008bridging}
\bibinfo{author}{Ide, S.}, \bibinfo{author}{Imanishi, K.},
  \bibinfo{author}{Yoshida, Y.}, \bibinfo{author}{Beroza, G.~C.} \&
  \bibinfo{author}{Shelly, D.~R.}
\newblock \bibinfo{title}{Bridging the gap between seismically and geodetically
  detected slow earthquakes}.
\newblock \emph{\bibinfo{journal}{Geophysical Research Letters}}
  \textbf{\bibinfo{volume}{35}} (\bibinfo{year}{2008}).

\bibitem{bletery2020slip}
\bibinfo{author}{Bletery, Q.} \& \bibinfo{author}{Nocquet, J.-M.}
\newblock \bibinfo{title}{{Slip bursts during coalescence of slow slip events
  in Cascadia}}.
\newblock \emph{\bibinfo{journal}{Nature communications}}
  \textbf{\bibinfo{volume}{11}}, \bibinfo{pages}{2159} (\bibinfo{year}{2020}).

\bibitem{heki1997silent}
\bibinfo{author}{Heki, K.}, \bibinfo{author}{Miyazaki, S.} \&
  \bibinfo{author}{Tsuji, H.}
\newblock \bibinfo{title}{{Silent fault slip following an interplate thrust
  earthquake at the Japan Trench}}.
\newblock \emph{\bibinfo{journal}{Nature}} \textbf{\bibinfo{volume}{386}},
  \bibinfo{pages}{595--598} (\bibinfo{year}{1997}).

\bibitem{ozawa2012preceding}
\bibinfo{author}{Ozawa, S.} \emph{et~al.}
\newblock \bibinfo{title}{{Preceding, coseismic, and postseismic slips of the
  2011 Tohoku earthquake, Japan}}.
\newblock \emph{\bibinfo{journal}{Journal of Geophysical Research: Solid
  Earth}} \textbf{\bibinfo{volume}{117}} (\bibinfo{year}{2012}).

\bibitem{churchill2022afterslip}
\bibinfo{author}{Churchill, R.}, \bibinfo{author}{Werner, M.},
  \bibinfo{author}{Biggs, J.} \& \bibinfo{author}{Fagereng, {\AA}.}
\newblock \bibinfo{title}{Afterslip moment scaling and variability from a
  global compilation of estimates}.
\newblock \emph{\bibinfo{journal}{Journal of Geophysical Research: Solid
  Earth}} \textbf{\bibinfo{volume}{127}}, \bibinfo{pages}{e2021JB023897}
  (\bibinfo{year}{2022}).

\bibitem{ozawa2024time}
\bibinfo{author}{Ozawa, S.}, \bibinfo{author}{Muneakane, H.} \&
  \bibinfo{author}{Suito, H.}
\newblock \bibinfo{title}{{Time-dependent modeling of slow-slip events along
  the Nankai Trough subduction zone, Japan, within the 2018--2023 period}}.
\newblock \emph{\bibinfo{journal}{Earth, Planets and Space}}
  \textbf{\bibinfo{volume}{76}}, \bibinfo{pages}{23} (\bibinfo{year}{2024}).

\bibitem{ohnaka1992earthquake}
\bibinfo{author}{Ohnaka, M.}
\newblock \bibinfo{title}{Earthquake source nucleation: a physical model for
  short-term precursors}.
\newblock \emph{\bibinfo{journal}{Tectonophysics}}
  \textbf{\bibinfo{volume}{211}}, \bibinfo{pages}{149--178}
  (\bibinfo{year}{1992}).

\bibitem{liu2007spontaneous}
\bibinfo{author}{Liu, Y.} \& \bibinfo{author}{Rice, J.~R.}
\newblock \bibinfo{title}{Spontaneous and triggered aseismic deformation
  transients in a subduction fault model}.
\newblock \emph{\bibinfo{journal}{Journal of Geophysical Research: Solid
  Earth}} \textbf{\bibinfo{volume}{112}} (\bibinfo{year}{2007}).

\bibitem{noda2013large}
\bibinfo{author}{Noda, H.}, \bibinfo{author}{Nakatani, M.} \&
  \bibinfo{author}{Hori, T.}
\newblock \bibinfo{title}{{Large nucleation before large earthquakes is
  sometimes skipped due to cascade-up—Implications from a rate and state
  simulation of faults with hierarchical asperities}}.
\newblock \emph{\bibinfo{journal}{Journal of Geophysical Research: Solid
  Earth}} \textbf{\bibinfo{volume}{118}}, \bibinfo{pages}{2924--2952}
  (\bibinfo{year}{2013}).

\bibitem{rubin2008episodic}
\bibinfo{author}{Rubin, A.~M.}
\newblock \bibinfo{title}{Episodic slow slip events and rate-and-state
  friction}.
\newblock \emph{\bibinfo{journal}{Journal of Geophysical Research: Solid
  Earth}} \textbf{\bibinfo{volume}{113}} (\bibinfo{year}{2008}).

\bibitem{liu2009slow}
\bibinfo{author}{Liu, Y.} \& \bibinfo{author}{Rice, J.~R.}
\newblock \bibinfo{title}{{Slow slip predictions based on granite and gabbro
  friction data compared to GPS measurements in northern Cascadia}}.
\newblock \emph{\bibinfo{journal}{Journal of Geophysical Research: Solid
  Earth}} \textbf{\bibinfo{volume}{114}} (\bibinfo{year}{2009}).

\bibitem{truttmann2024slow}
\bibinfo{author}{Truttmann, S.}, \bibinfo{author}{Poulet, T.},
  \bibinfo{author}{Wallace, L.}, \bibinfo{author}{Herwegh, M.} \&
  \bibinfo{author}{Veveakis, M.}
\newblock \bibinfo{title}{{Slow slip events in New Zealand: irregular, yet
  predictable?}}
\newblock \emph{\bibinfo{journal}{Geophysical Research Letters}}
  \textbf{\bibinfo{volume}{51}}, \bibinfo{pages}{e2023GL107741}
  (\bibinfo{year}{2024}).

\bibitem{kaveh2025spatiotemporal}
\bibinfo{author}{Kaveh, H.}, \bibinfo{author}{Avouac, J.~P.} \&
  \bibinfo{author}{Stuart, A.~M.}
\newblock \bibinfo{title}{Spatiotemporal forecast of extreme events in a
  chaotic model of slow slip events}.
\newblock \emph{\bibinfo{journal}{Geophysical Journal International}}
  \textbf{\bibinfo{volume}{240}}, \bibinfo{pages}{870--885}
  (\bibinfo{year}{2025}).

\bibitem{sherrill2021new}
\bibinfo{author}{Sherrill, E.} \& \bibinfo{author}{Johnson, K.}
\newblock \bibinfo{title}{{New insights into the slip budget at Nankai: An
  iterative approach to estimate coseismic slip and afterslip}}.
\newblock \emph{\bibinfo{journal}{Journal of Geophysical Research: Solid
  Earth}} \textbf{\bibinfo{volume}{126}}, \bibinfo{pages}{2020JB020833}
  (\bibinfo{year}{2021}).

\bibitem{okada1985surface}
\bibinfo{author}{Okada, Y.}
\newblock \bibinfo{title}{Surface deformation due to shear and tensile faults
  in a half-space}.
\newblock \emph{\bibinfo{journal}{Bull. Seismol. Soc. Am}}
  \textbf{\bibinfo{volume}{75}}, \bibinfo{pages}{1135--1154}
  (\bibinfo{year}{1985}).

\bibitem{zou2005regularization}
\bibinfo{author}{Zou, H.} \& \bibinfo{author}{Hastie, T.}
\newblock \bibinfo{title}{Regularization and variable selection via the elastic
  net}.
\newblock \emph{\bibinfo{journal}{Journal of the Royal Statistical Society
  Series B: Statistical Methodology}} \textbf{\bibinfo{volume}{67}},
  \bibinfo{pages}{301--320} (\bibinfo{year}{2005}).

\bibitem{geonet_doi}
\bibinfo{author}{{Geodetic Observation Center, Geospatial Information Authority
  of Japan}}.
\newblock \bibinfo{title}{{GNSS data}} (\bibinfo{year}{1996$\sim$}).

\bibitem{tanaka_2024_14242258}
\bibinfo{author}{Tanaka, Y.} \& \bibinfo{author}{Nishimura, T.}
\newblock \bibinfo{title}{{Input GNSS time series data for Tanaka et al.
  (2024), JGR Solid Earth }} (\bibinfo{year}{2024}).
\newblock \urlprefix\url{https://doi.org/10.5281/zenodo.14242258}.

\bibitem{kano2018development}
\bibinfo{author}{Kano, M.} \emph{et~al.}
\newblock \bibinfo{title}{Development of a slow earthquake database}.
\newblock \emph{\bibinfo{journal}{Seismological Research Letters}}
  \textbf{\bibinfo{volume}{89}}, \bibinfo{pages}{1566--1575}
  (\bibinfo{year}{2018}).

\end{thebibliography}

\bibliographystyle{naturemag}

\section*{Acknowledgements}
We thank Ryoko Nakata for her advice on data-catalog curation and Yuji Itoh and Rikuto Fukushima for their valuable comments on the source analysis of SSEs. This study was supported by JSPS KAKENHI Grants (23K19082 and 21H05206).

\section*{Author contributions}
D.S. performed the statistical analysis and synthesized the mechanical interpretation. T.H. surveyed the data and conceptualized the research question. T.I. supervised the geodetic inverse analysis. M.K. provided the draft code of inverse analysis. Y.T. processed the original GNSS data. All the authors contributed to discussion and manuscript writing.

\section*{Competing interests}
There are no competing interests to declare.

\section*{Data availability}
The GNSS data used in this study were downloaded from the GNSS Earth Observation Network System (GEONET, https://doi.org/10.57499/GSI\_GNSS\_2025\_001) operated by the Geospatial Information Authority of Japan~\cite{geonet_doi}, processed by courtesy of Dr. Takuya Nishimura~\cite{tanaka_2024_14242258}. The tremor catalog was downloaded from the Slow Earthquake Database~[http://www-solid.eps.s.u-tokyo.ac.jp/{\texttildelow}sloweq \cite{kano2018development}].

\section*{Additional information}
\paragraph*{Supplementary information}  The online version contains the supplementary material consisting of Figs.~S1 and S2 and Tables~S1--S3
available at https(...the address assigned by the journal...).
\paragraph*{Correspondence} Correspondence and requests for materials should be addressed to D.S.

\newpage
\pagenumbering{gobble}

\selectlanguage{english}
\FloatBarrier


\subsection*{Supplementary materials}
Figs. S1 and S2\\
Tables S1 to S3


\newpage


\renewcommand{\thefigure}{S\arabic{figure}}
\renewcommand{\thetable}{S\arabic{table}}
\renewcommand{\theequation}{S\arabic{equation}}
\renewcommand{\thepage}{S\arabic{page}}
\setcounter{figure}{0}
\setcounter{table}{0}
\setcounter{equation}{0}
\setcounter{page}{1} 

\section*{Supplementary Materials for Postseismicity of slow-slip doublets discerned on the outermost of the Nankai Trough subduction megathrust}

	Dye~SK~Sato$^{\ast}$,
	Takane~Hori,
	Takeshi~Iinuma,
	Masayuki~Kano,
	Yusuke~Tanaka
\\ 
\small$^\ast$Corresponding author. Email: daisukes@jamstec.go.jp\\

\subsubsection*{This PDF file includes:}
Figures S1 and S2\\
Tables S1 to S3\\

\newpage







\begin{figure} 
	\centering
    \includegraphics[width=0.6\textwidth]{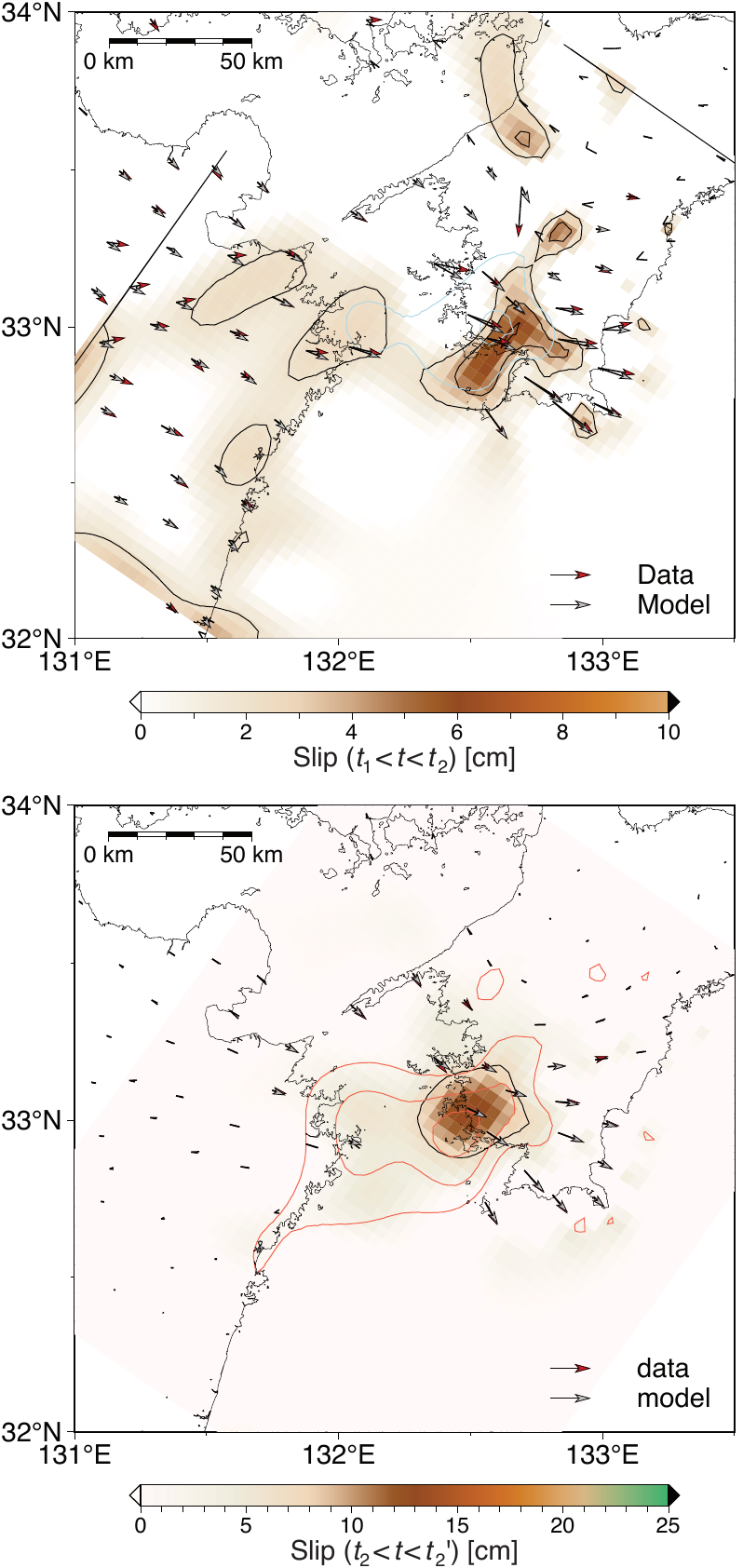} 

	\caption{\textbf{Slip estimates for $t_1<t<t_2$ and $t_2<t<t_2^\prime$.}
		 The most probable estimates using the prior constraint of the Elastic Net are evaluated for $t_1<t<t_2$ (top) and $t_2<t<t_2^\prime$ (bottom).}
	\label{fig:sup1} 
\end{figure}

\begin{figure} 
	\centering
    \includegraphics[width=1\textwidth]{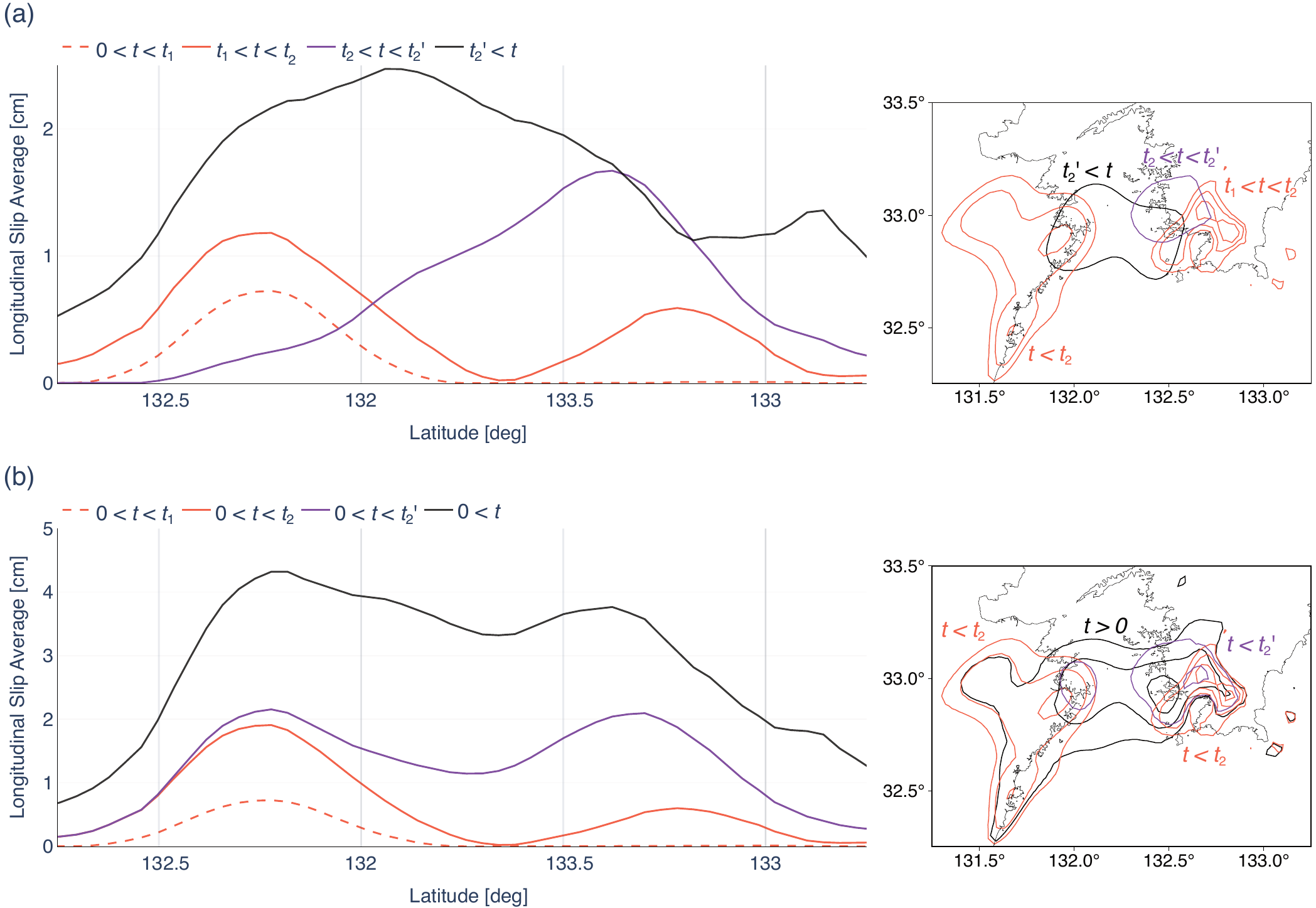} 

	\caption{Differential and cumulative slips calculated from slip estimates at different time windows. Longitudinal averages are calculated as in Fig. 3b but using the slip estimates for $t<t_1$ (Fig. 2c1), $t<t_2$ (Fig. 2b1), $t_2<t<t_2^\prime$ (Fig. S1, bottom), and $t>t_2^\prime$ (Fig. 2c3). (a) Differential slips. Longitudinal averages and contours are shown. Each slip estimate is plotted as is, but for $t_1<t<t_2$, where the slip is calculated from the difference in the estimated slip between $t<t_1$ and $t<t_2$. (b) Cumulative slips, calculated as the sums of differential slips shown in Fig.~\ref{fig:sup2}a.}
	\label{fig:sup2} 
\end{figure}


\begin{table} 
	\centering
	\caption{\textbf{Estimates of linear-plus-log fits.}
		Least-square estimates of $t_1$, $t_2$, and $t_{\rm c}$ are listed for the case using an empirical function $F_1$ (eq.~1). The estimates are shown for respective stacked data of the postseismicwise group (Post), post-\&co-seismicwise group (Co\&Post), and total stations (Total). The error terms express variances in the least squares.}
	\label{tab:sup1} 
	\begin{tabular}{c|ccc} 
		\hline
		$F_1$ fit (eq.~1) & Post & Co\&Post & Total \\
		\hline
		$t_1$ [date] & 2009-08-11$\pm$4 & 2009-07-21$\pm$3 & 2009-08-16$\pm$5 \\
		$t_2$ [date] & 2010-01-29$\pm$1 & 2010-02-03$\pm$1 & 2010-01-13$\pm$5 \\
		$t_{\rm c}$ [days] & 85$\pm$3 & 43$\pm$2 & 337$\pm$27 \\
	\end{tabular}
\end{table}

\begin{table} 
	\centering
	\caption{\textbf{Modified estimates of linear-plus-log fits using a truncated time window excluding the last half a year.}
		The list of the same estimates as Table~\ref{tab:sup1} but for the truncated time window [2006-07-01, 2010-12-01], which excludes the last 182 days of the original period [2006-07-01, 2011-06-01] during which the Central Shikoku SSE is active.}
	\label{tab:sup2} 
	\begin{tabular}{c|ccc} 
		\hline
		$F_1$ fit (eq.~1, truncated) & Post & Co\&Post & Total \\
		\hline
		$t_1$ [date] & 2009-08-11$\pm$4 & 2009-07-21$\pm$3 & 2009-08-13$\pm$5 \\
		$t_2$ [date] & 2010-02-04$\pm$1 & 2010-02-02$\pm$1 & 2010-02-01$\pm$3 \\
		$t_{\rm c}$ [days] & 51$\pm$3 & 48$\pm$3 & 56$\pm$6 \\
	\end{tabular}
\end{table}

\begin{table} 
	\centering
	\caption{\textbf{Estimates of exp-plus-log fits.}
		Least-square estimates of $t_2$ and $t_{\rm c}$ are listed for the case using an empirical function $F_2$ (eq.~2). 
        The error terms express variances in the least squares.
        The time $t_2^\prime$ where the first and second terms of $F_2$ take the same nonzero value is also evaluated. }
	\label{tab:sup3} 

	\begin{tabular}{c|ccc} 
		\hline
		$F_2$ fit (eq.~2) & Post & Co\&Post & Total \\
		\hline
		$t_2$ [date] & 2010-01-27$\pm$1 & 2010-02-01$\pm$1 & 2010-01-23$\pm$5 \\
		$t_{\rm c}$ [days] & 87$\pm$5 & 46$\pm$2 & 593$\pm$152\\
		$t_2^\prime$ [date] & 2010-02-26 & 2010-03-11 & 2010-05-27\\
	\end{tabular}
\end{table}
\end{document}